\documentclass[aps,prd,twocolumn,groupedaddress,showpacs]{revtex4}
\usepackage{graphicx,epsf,amssymb,amsmath,latexsym}
\newcommand{\be}{\begin{eqnarray}}
\newcommand{\ee}{\end{eqnarray}}

\newcommand{\expec}[1]{\mbox{$\langle\, #1\,\rangle$}}

\newcommand{\dd}{{\rm d}}
\begin{document}
\title{Holography and trace anomaly: what is the fate of
(brane-world) black holes?}
\author{Roberto Casadio}
\email{casadio@bo.infn.it}
\affiliation{Dipartimento di Fisica, Universit\`a di Bologna,
and I.N.F.N., Sezione di Bologna,
via Irnerio 46, 40126 Bologna, Italy}
\begin{abstract}
The holographic principle relates (classical) gravitational
waves in the {\em bulk\/} to quantum fluctuations and the Weyl anomaly
of a conformal field theory on the boundary (the {\em brane\/}).
One can thus argue that linear perturbations in the bulk of static
black holes located on the brane be related to the Hawking flux and
that (brane-world) black holes are therefore unstable.
We try to gain some information on such instability from established
knowledge of the Hawking radiation on the brane.
In this context, the well-known trace anomaly is used as a measure of
both the validity of the holographic picture and of the instability for
several proposed static brane metrics.
In light of the above analysis, we finally consider a time-dependent
metric as the (approximate) representation of the late stage of
evaporating black holes which is characterized by decreasing Hawking
temperature, in qualitative agreement with what is required by energy
conservation.
\end{abstract}
\pacs{04.70.-s, 04.70.Bw, 04.50.+h}
\maketitle
\section{Introduction}
\label{intro}
The holographic principle \cite{holography}, in the form of
the AdS-CFT conjecture \cite{AdSCFT}, when applied to the
Randall-Sundrum (RS) brane-world \cite{RS}, yields a relation
between classical gravitational perturbations in the
bulk and quantum fluctuations of conformal matter fields on
the brane \cite{soda}.
It was then proposed in Refs.~\cite{tanaka1,fabbri} that black hole
metrics which solve the bulk equations with brane boundary conditions,
and whose central singularities are located on the brane, genuinely
correspond to quantum corrected (semiclassical) black holes
on the brane, rather than to classical ones.
A major consequence of such a conjecture would be that no static
black holes exist in the brane-world \cite{bruni}, since
semiclassical corrections (approximately described by a conformal
field theory -- CFT -- on the brane) have the form of a flux
of energy from the source.
That black holes are unstable was already well known in the
four-dimensional theory, since the Hawking effect \cite{hawking}
makes such objects evaporate, and the semiclassical Einstein
equations should hence contain the {\em back-reaction\/} of the
evaporation flux on the metric.
The novelty is that, from the bulk side (of the AdS-CFT
correspondence), one understands the Hawking radiation in terms
of classical gravitational waves whose origin is the acceleration
the black holes are subject to, living on a non-geodesic
hypersurface of the (asymptotically) anti-de Sitter space-time
AdS$_5$~\footnote{This argument would suggest that the
brane-world is intrinsically unstable under the addition of
any matter content, since matter on the brane does not follow
geodesics of the five-dimensional metric and is accelerated.
From the phenomenological point of view, it is then important
to establish the typical times for such an instability to
become observable.}.
\par
In Ref.~\cite{cm} a method was proposed by which static brane
metrics, such as the asymptotically flat ones put forward in
Refs.~\cite{maartens,germani,kar,cfm}, can be extended into the
bulk~\footnote{For more general brane solutions see
Refs.~\cite{visser} and for a nice review of black holes and
extra dimensions see Ref.~\cite{cavaglia}.}.
The method makes use of the multipole expansion (with respect to
the usual areal radial coordinate $r$ on the brane) and proved
particularly well suited for very large black holes with
\be
M\gg \sigma^{-1}
\ ,
\label{large}
\ee
where $M$ is the four-dimensional
Arnowitt-Deser-Misner (ADM) mass parameter (in geometrical units)
and $3\,\sigma$ the brane tension (times the five-dimensional
gravitational constant).
The main result is that the horizon of an astrophysical size black
hole has the shape of a ``pancake'' (see also Ref.~\cite{katz})
and its area is roughly equal to the four-dimensional expression
(in terms of $M$).
To the extent at which the employed approximation is reliable,
no singular behavior in the curvature and Kretschmann scalars in
the bulk was found, contrary to the case of the Black String
(BS) of Ref.~\cite{chamblin}.
Possible caveats of this approach have already been thoroughly
discussed in Ref.~\cite{cm}.
In particular, the convergence of the multiple expansion on the
axis of cylindrical symmetry which extends into the bulk is hard
to test and the resulting metrics are not completely reliable
thereon.
As a consequence, it is hard to determine whether the bulk geometry
contains singularities on the axis (while no singularity seems
to occur far from it) and whether it is asymptotically AdS away
from the brane near the axis (while it appears asymptotically AdS
far from the axis).
In fact, in recent numerical works, regular brane metrics were
shown to develop singularities in the bulk when extended by a
different method \cite{shinkai} or problems emerged when trying
to describe large black holes in asymptotically AdS bulk
\cite{kudoh}.
\par
If the conjecture of Refs.~\cite{tanaka1,fabbri,bruni} is correct,
it then follows that the static bulk solutions found in
Ref.~\cite{cm} have singularities on the cylindrical axis
(possibly far away from the brane) or, at least, are unstable
under linear perturbations (of the metric in the bulk), as it
occurs for the BS \cite{gregory}.
In either case, it is likely that such metrics will evolve toward
different (more stable but yet unknown) configurations \cite{bruni}.
Since the metric elements in Ref.~\cite{cm} are expressed as
sums of many terms (multipoles) and those in
Refs.~\cite{shinkai,kudoh} are expressed only numerically,
it is practically impossible to carry out a linear perturbation
analysis on such backgrounds.
One could otherwise try to use the AdS-CFT correspondence backwards
in order to estimate the overall effect of bulk perturbations
from the established knowledge of the Hawking radiation on the
brane.
Some information on the latter can be determined straightforwardly
from standard four-dimensional expressions provided the brane
metric is given (see, e.g., Ref.~\cite{birrell}).
One must then check that such information is consistent with
known features of the AdS-CFT correspondence before drawing any
conclusion about the bulk stability.
\par
In fact, the AdS-CFT correspondence requires some general
conditions to hold.
First of all, one needs the planar limit of the large ${\mathcal N}$
expansion of the CFT, that is a large number of conformal fields
\cite{AdSCFT}
\be
{\mathcal N}\sim \sigma^{-2}\,\ell_{\rm p}^{-2}\gg 1
\ .
\label{sig}
\ee
where, in RS models, the four-dimensional Newton constant
$8\,\pi\,G_N=\ell_{\rm p}^2$ ($\ell_{\rm p}$ being the Planck length)
is related to the fundamental gravitational length $\ell_{\rm g}$ in
five dimensions by $\ell_{\rm p}^2=\sigma\,\ell_{\rm g}^3$
\cite{RS,shiromizu}.
Eq.~(\ref{sig}) is therefore tantamount to $\ell_{\rm g}\gg\ell_{\rm p}$
and assures that stringy effects are negligible.
Moreover, the presence of the brane introduces a normalizable 
four-dimensional graviton and an ultra-violet (UV) cut-off in
the CFT,
\be
\lambda_{\rm UV}\sim \sigma^{-1}
\ .
\label{UV}
\ee
The latter must of course be much shorter than any physical low energy
scale $\lambda_{\rm IR}$ of interest,
\be
\lambda_{\rm UV}\ll\lambda_{\rm IR}
\ ,
\label{cut-offs}
\ee
in order for the CFT description of the brane-world to be
consistent.
A further remark is in order about the relevance of the
bulk being asymptotically AdS away from the brane.
We have already mentioned that, according to Ref.~\cite{cm},
possible singularities in the bulk (and the corresponding
deviations from asymptotic AdS) should be located on the axis.
To what extent such singularities restrict the holographic
description is hard to tell {\em a priori\/} and
a general criterion for the validity of the AdS-CFT
correspondence will therefore be given in
Section~\ref{general} [see Eq.~(\ref{GammaCFT}) and
the following discussion]. 
\par
So far, no practical advantage seems to emerge from the
holographic description over the standard four-dimensional
treatment of the back-reaction problem.
There is however a point of attack: As we shall review shortly,
in the brane-world a vacuum (brane) solution needs to satisfy
just one equation, whose analysis is therefore significantly
simpler than the full set of four-dimensional vacuum Einstein
equations.
This will allow us to attempt an approximate description of the
late stage of the evaporation which is in qualitative agreement
with earlier studies of black holes as extended objects in the
microcanonical picture \cite{harms,mfd}, and is characterized by
a black hole temperature which vanishes along with the mass of
the black hole.
Let us remark that the total energy of the system (black hole plus
Hawking radiation) is conserved in the microcanonical ensemble and
this suggests that the final evolution of a black hole is really
unitary.
\par
In the next Section we summarize the general concepts with a
particular emphasis on the trace anomaly as derived from the
point of view of four-dimensional quantum field theory and
its comparison with that predicted by the AdS-CFT correspondence.
In Section~\ref{static}, we then apply this formalism to candidate
static black holes in order to check the reliability of their
holographic picture and stability.
Our conclusions first of all support the view given in
Refs.~\cite{cm,wiseman} that static metrics are a good
approximation for astrophysical black holes.
Moreover, some of the brane metrics analyzed in Ref.~\cite{cm,cfm}
are shown to allow for a closer holographic interpretation
and to be more stable than the BS.
This suggests that brane-world black holes might evaporate
more slowly than they would do in a four-dimensional space-time
already for very large ADM masses~\footnote{On considering the
very large number of conformal fields required by the AdS-CFT
correspondence, the authors of Ref.~\cite{garcia} however come
to the opposite conclusion.}.
Finally, in Section~\ref{time} we discuss a possible candidate
time-dependent metric to estimate the late stage of the evaporation
by self-consistently including the trace anomaly in the relevant
vacuum equation.
\par
We shall adopt the brane metric signature $(-,+,+,+)$ and units with
$\hbar=c=1$.
Latin indices $i,j,\ldots$ will denote brane coordinates throughout
the paper.
\section{General framework}
\label{general}
The five-dimensional Einstein equations in (asymptotically) AdS$_5$
with bulk cosmological constant $\Lambda$ can be projected onto the
brane by making use of the Gauss-Codazzi relations and Israel's junction
conditions (see Ref.~\cite{shiromizu} for the details).
For the RS case which we consider throughout the paper
$\Lambda=-\sigma^2\,\ell_{\rm g}^3/6$ \cite{RS}, so that the brane
cosmological constant vanishes and the effective four-dimensional
Einstein equations become
\be
G_{ij}=\ell_{\rm p}^2\,\tau_{ij}+{\ell_{\rm p}^4\over\sigma^2}\,\pi_{ij}+E_{ij}
\ ,
\ee
where $G_{ij}=R_{ij}-(1/2)\,R\,g_{ij}$ is the four-dimensional Einstein
tensor, $\tau_{ij}$ is the energy-momentum tensor of matter localized
on the brane (there is no matter in the bulk) and
\be
\pi_{ij}=-{1\over 4}\,\tau_{ik}\,\tau_j^{\ k}
+{1\over 12}\,\tau\,\tau_{ij}
+{1\over 8}\,g_{ij}\,\tau_{kl}\,\tau{kl}
-{1\over 24}\,g_{ij}\,\tau^2
\ .
\ee
Where no matter appears on the brane ($\tau_{ij}=0$), the existence of an
extra spatial dimension manifests itself in the brane-world only in the
form of the non-local source term $E_{ij}$, which is (minus) the
projection of the bulk Weyl tensor on the brane and must be traceless
\cite{shiromizu,tanaka1}.
Vacuum brane metrics therefore satisfy
\be
&R_{ij}=E_{ij}&
\\
&\Downarrow
\nonumber
\\
&R=0
\ .
&
\label{R}
\ee
Of course, Eq.~(\ref{R}) is a weaker condition than the
four-dimensional vacuum equation $R_{ij}=0$ and, consequently,
Birkhoff's theorem does not necessarily hold for spherically
symmetric vacuum brane metrics.
\par
The AdS-CFT correspondence should relate the tensor $E_{ij}$ representing
(classical) gravitational waves in the bulk to the expectation value of
the (renormalized) energy-momentum tensor of conformal fields on the brane
\cite{soda}.
Let us denote the latter by $\expec{T_{ij}}$.
One should then have
\be
E^i_{\ j}\sim\ell_{\rm p}^2\,\expec{T^i_{\ j}}
\ .
\ee
Since the left hand side above is traceless, such a correspondence can
hold if $\expec{T}\equiv\expec{T^i_{\ i}}=0$, that is, if the conformal
symmetry is not anomalous.
Of course, this requires that the UV cut-off (\ref{UV}) be much shorter
than any physical length scale in the system.
It also requires a ``flat'' brane (i.e., the absence of any intrinsic
four-dimensional length associated with the background) otherwise
the CFT will also be affected by that scale.
For a black hole, the gravitational radius $r_{\rm h}\sim M$ is a natural
length scale and one therefore expects that only CFT modes with
wavelengths much shorter than $r_{\rm h}$~\footnote{As is well known,
this is also the typical wavelength of the Hawking radiation
\cite{hawking} and one naively perceives a connection between the
two effects.} (and still much larger than $\sigma^{-1}$) propagate
freely.
One then finds that the necessary condition (\ref{cut-offs}) is
equivalent to Eq.~(\ref{large}), that is a reliable CFT descritpion
of the Hawking radiation might be possible only for very large black
holes of the kind considered in Ref.~\cite{cm}.
\par
From the point of view of the AdS-CFT correspondence, it is the
value of bulk perturbations at the boundary that acts as a source for
the CFT fields and can give rise to $\expec{T}_{\rm CFT}\not=0$.
As a check, one can compare with the trace anomaly induced by the presence
of a brane as a boundary of AdS in several theories in which the AdS-CFT
applies.
Since we are just interested in a four-dimensional brane, the case
of relevance to us is that of (${\mathcal N}$ stacked) D3-branes
(possibly with $R=0$) embedded in AdS$_5$.
For such a configuration one finds the holographic Weyl anomaly
\cite{skenderis}
\be
\expec{T}_{\rm CFT}=
{1\over 4\,\ell_{\rm p}^2\,\sigma^2}\,
\left(R_{ij}\,R^{ij}-{1\over 3}\,R^2\right)
\ ,
\label{Tcft}
\ee
which reproduces the conformal anomaly of the four-dimensional
${\mathcal M}=4$ superconformal $SU({\mathcal N})$ gauge theory
in the large ${\mathcal N}$ limit (\ref{sig}) and vanishes in a
four-dimensional (Ricci flat) vacuum~\footnote{Notably, the Weyl
anomaly evaluated in the Euclidean AdS$_5$ (in which the brane-world
is a four-sphere) precisely yields the brane tension $\sigma$
\cite{hawk} (see also Refs.~\cite{odintsov}).
For the connection between the Euclidean and the Lorentzian
versions of the AdS-CFT correspondence, see
Refs.~\cite{AdSCFT,sato}}.
\par
On the other hand, the trace anomaly of the pertinent 
four-dimensional field theory, $\expec{T}_{\rm 4D}=\expec{T^i_{\ i}}$,
can be evaluated independently.
It is given in terms of geometrical quantities as well and 
numerical coefficients which depend on the matter fields.
Further, it does not usually vanish on a curved background
(even if it is Ricci flat) because, contrary to $\expec{T}_{\rm CFT}$,
it also contains the Kretschmann scalar $R_{ijkl}\,R^{ijkl}$.
For example, one finds for $n_{\rm B}$ boson fields
(see, e.g.~Ref.~\cite{birrell})
\be
\expec{T}_{\rm 4D}=
{n_{\rm B}\over 2880\,\pi^2}\,
\left(R_{ijkl}\,R^{ijkl}-R_{ij}\,R^{ij}-\Box R\right)
\ .
\ee
The term $\Box R$, which is renormalization dependent, would however
vanish according to Eq.~(\ref{R}) but we include such term for future
reference (see, in particular, Section~\ref{time}).
It is this non-vanishing trace $\expec{T}_{\rm 4D}$ which measures
the actual violation of the conformal symmetry on the brane.
\par
If $\expec{T}_{\rm 4D}\not=\expec{T}_{\rm CFT}$, one needs more
than the AdS-CFT correspondence to describe the brane physics for
the chosen background.
In other words, this inequality can be interpreted as signaling the
excitation of other matter fields living on the brane (with
$\tau\equiv\tau^i_{\ i}\sim\expec{T}_{\rm 4D}-\expec{T}_{\rm CFT}$).
The relative difference with respect to
$\expec{T}_{\rm CFT}$,
\be
\Gamma_{\rm CFT}\equiv
\left|{\expec{T}_{\rm 4D}-\expec{T}_{\rm CFT}\over
\expec{T}_{\rm CFT}}\right|
\ ,
\label{GammaCFT}
\ee
can then be used to estimate to what extent classical gravitational
waves in the bulk determine matter fluctuations on the
brane~\footnote{One should be very cautious in applying this
argument near the horizon, where the role of (trans)-Planckian
physics is not clear (see e.g. \cite{c1} and References therein).
In the following a (sufficiently) large $r$ expansion will always
be assumed.}.
If $\Gamma_{\rm CFT}\ll 1$, then the AdS-CFT conjecture implies that
the  (quantum) brane and (classical) bulk descriptions of black holes
are equivalent.
Otherwise, since the holography can just account for that part of
the Hawking flux which is responsible for $\expec{T}_{\rm CFT}$,
the ratio $\Gamma_{\rm CFT}$ is also a measure of the relative
strength of brane fluctuations (involving other matter modes)
with respect to bulk gravitational waves.
\par
From the four-dimensional point of view, the trace anomaly is
evidence that one is quantizing matter fields, by means of
the background field method, on the ``wrong'' (i.e.~unstable)
background metric.
One should instead find a background and matter state 
(both necessarily time-dependent) whose metric and energy-momentum
tensor solve all relevant field equations simultaneously.
This is the aforementioned back-reaction problem of Hawking
radiation, whose solution is still out of grasp after several
decades from the discovery of black hole evaporation.
The authors of Ref.~\cite{fabbri} argue that, because of the
AdS-CFT correspondence, the problem of describing a brane-world
black hole is just as difficult as the (four-dimensional)
back-reaction problem itself.
One could go even further and claim that it is {\em at least\/}
as difficult, since for an holographic black hole the AdS-CFT
should be exact and $\expec{T}_{\rm 4D}=\expec{T}_{\rm CFT}$
(i.e., $\Gamma_{\rm CFT}=0$), but realistic black holes might
involve more ``ingredients''.
If however one focuses on the brane description, and just considers
Eq.~(\ref{R}), the task will simplify considerably.
\section{Static black holes?}
\label{static}
We first want to analyze both the semiclassical stability and
holography of candidate static brane-world black holes.
They are described by asymptotically flat, spherically symmetric
metrics which solve Eq.~(\ref{R}), and can be put in the form
\be
\dd s^2=-N(r)\,\dd t^2+A(r)\,\dd r^2+r^2\,\dd \Omega^2
\ ,
\label{g}
\ee
where $\dd\Omega^2=\dd\theta^2+\sin^2\theta\,\dd\phi^2$, and for the
functions $N$ and $A$ we shall now consider several cases
previously appeared in the literature.
\subsection{4D and CFT trace anomalies}
The first step is to compute $\expec{T^i_{\ j}}$,
$\expec{T}_{\rm 4D}$ and $\expec{T}_{\rm CFT}$.
The ratio $\Gamma_{\rm CFT}$ will then give a measure
of the reliability of the holographic picture.
\subsubsection{Black String}
We take the Schwarzschild line element (corresponding to the BS
\cite{chamblin}),
\be
N={1\over A}=1-{2\,M\over r}
\ ,
\label{g0}
\ee
as the reference brane metric, whose horizon radius is $r_{\rm h}=2\,M$.
The trace of the energy-momentum tensor for $n_{\rm B}$ boson fields
on this background is
\be
^{\rm S}\expec{T}_{\rm 4D}={n_{\rm B}\,M^2\over 60\,\pi^2\,r^6}
\ ,
\label{TS}
\ee
and four-dimensional covariant conservation relates it to those parts
of the diagonal components of the energy-momentum tensor which do
not depend on the quantum state of the radiation
\cite{christensen}.
Moreover, in the Unruh vacuum, one also has the flux of outgoing
Hawking radiation \cite{birrell,christensen},
\be
^{\rm S}\expec{T^t_{\ t}}=
\,^{\rm S}\expec{T^t_{\ r}}=
-\,^{\rm S}\expec{T^r_{\ t}}=
-\,^{\rm S}\expec{T^r_{\ r}}
\sim{n_{\rm B}\,K\over M^2\,r^2}
\ ,
\label{T_S}
\ee
where $K$ is a dimensionless constant.
It is therefore expected that the components of the Ricci tensor
of the metric which incorporates the back-reaction of the Hawking
radiation have a leading behavior
\be
^{\rm S}E^i_{\ j}\sim{n_{\rm B}\,\ell_{\rm p}^2\over M^2\,r^2}
\ .
\label{RicciS}
\ee
The above terms, representing a steady flux of radiation, are
consistent within the adiabatic approximation for which the
background (brane) metric is held static.
In fact, the constant $K$ (formally) arises from an integration
performed over an infinite interval of (Euclidean) time, so as
to include all the poles in the Wightman function of the radiation
field which yield the Planckian spectrum (the integral diverges and
is divided by the length of the time interval to estimate the
average flux per unit time \cite{hawking}).
During this large time, the change in the ADM mass is neglected.
In order to solve for the back-reaction, one should instead consider
the full time dependence of the metric and black hole source.
\par
From the point of view of the bulk space-time, the Hawking flux
just modifies the junction equations \cite{israel} at the brane.
The latter, on account of the orbifold symmetry Z$_2$, will in general
read \cite{shiromizu}
\be
\left[K_{ij}\right]
=\ell_{\rm g}^3\,
\left(T_{ij}-{1\over 3}\,g_{ij}\,T\right)
\  .
\ee
The quantity $\left[K_{ij}\right]$ is the jump in the extrinsic
curvature of the brane and $T_{ij}$ the source term localized
on the brane which, for the case at hand, contains the vacuum
energy $3\,\sigma$ and the Hawking flux.
One thus has
\be
\left[K_{ij}\right]=-\sigma\,g_{ij}
+{\ell_{\rm p}^2\over\sigma}\,
\left(\,^{\rm S}\expec{T_{ij}}
-{1\over 3}\,g_{ij}\,^{\rm S}\expec{T}_{\rm 4D}\right)
\ .
\label{junction}
\ee
Since $r>2\,M$ outside the horizon, the second term in the
right hand side above is negligible with respect to the first
one everywhere in the space accessible to an external observer
if
\be
M\,\sigma\gg {\ell_{\rm p}\over M}
\ .
\label{cond}
\ee
This shows that the Hawking radiation just gives rise to a small
perturbation of the bulk metric for black holes of astrophysical
size [for which Eq.~(\ref{large}) holds].
This is assumed in the approach of Ref.~\cite{cm} (and also in
the numerical analysis of Refs.~\cite{wiseman}) to justify
staticity~\footnote{The presence of non-vanishing off-diagonal
components of the energy-momentum tensor (\ref{T_S}) while $K_{ij}$
and $G_{ij}$ are diagonal may seem to invalidate this argument.
However, let us recall that any comparisons between different
contributions should be better made in terms of scalar quantities,
such as $K^i_{\ i}$ and $K_{ij}\,K^{ij}$ on the one hand, and
$\expec{T}$ and $\expec{T_{ij}}\,\expec{T^{ij}}$ on the other.
In so doing, one precisely obtains conditions of the form given in
Eq.~(\ref{cond}) with numerical coefficients of order $1$.}.
\par
However, the holographic Weyl anomaly (\ref{Tcft}) vanishes
for this metric (the ratio $\Gamma_{\rm CFT}\to\infty$) since
$R_{ij}=0$, and drawing any conclusion from the AdS-CFT
correspondence looks questionable.
\subsubsection{Case I}
From Refs.~\cite{cfm,germani}, we consider the functions
\be
N=1-{2\,M\over r}
\ ,\ \ \
A={\left(1-{3\,M\over 2\,r}\right)\over
N\,\left[1-{3\,M\over 2\,r}\,(1+{4\over 9}\,\eta)\right]}
\ ,
\label{gI}
\ee
the causal structure of whose geometry was analyzed in details
in Ref.~\cite{cfm}.
One finds that the non-vanishing Ricci tensor components
are given by
\be
&&^{\rm I}R^t_{\ t}={4\,\eta\,M^2\over 3\,(2\,r-3\,M)^2\,r^2}
\nonumber
\\
&&^{\rm I}R^r_{\ r}= -{4\,\eta\,M\over 3\,(2\,r-3\,M)\,r^2}
\label{RicciI}
\\
&&^{\rm I}R^\theta_{\ \theta}
=\,^{\rm I}R^\phi_{\ \phi}
= -{4\,\eta\,M\,(r-M)\over 3\,(2\,r-3\,M)^2\,r^2}
\ .
\nonumber
\ee
Note that the corrections in Eq.~(\ref{RicciS}) that one obtains
from the Hawking radiation dominate (in the large $r$ approximation)
over those following from the components in Eq.~(\ref{RicciI}).
This is expected since the metric (\ref{gI}) does not contain
an outgoing flux and is asymptotically flat.
However, the corrections in Eq.~(\ref{RicciI}) certainly overcome
the Hawking flux in the interval
\be
1\ll {r\over M}\lesssim|\eta|\,{M^2\over\ell_{\rm p}^2}
\ ,
\label{int}
\ee
whose extension can be very large for astrophysical black
holes (with $M\gg\sigma^{-1}\gtrsim\ell_{\rm p}$) provided $\eta$
is not infinitesimal~\footnote{For the
typical solar mass $M\sim 1\,$km, and $\sigma^{-1}\lesssim 1\,$mm,
one has $M\sim 10^{38}\,\ell_{\rm p}\gtrsim 10^7\,\sigma^{-1}$ \cite{cm}.
This makes the upper limit in the interval (\ref{int}) of
the order of $10^{76}\,|\eta|\,$km or larger, which is practically
an infinite extension, even if one considers the experimental bound
$|\eta|\lesssim 10^{-4}$ \cite{will}.}.
Moreover, the trace of the energy-momentum tensor of a boson
field on this four-dimensional background has a leading behavior
for large $r$ given by (see Appendix~\ref{anomalies} for more details)
\be
^{\rm I}\expec{T}_{\rm 4D}\simeq
\left(1+{\eta\over 3}+{\eta^2\over 24}\right)\,
^{\rm S}\expec{T}_{\rm 4D}
\ ,
\label{TI}
\ee
which is of the same order in $1/r$ as the trace in Eq.~(\ref{TS}).
\par
The expected trace anomaly from the AdS-CFT correspondence on this
background is of the same order as $^{\rm I}\expec{T}_{\rm 4D}$,
namely
\be
^{\rm I}\expec{T}_{\rm CFT}\sim
{\eta^2\,M^2\over 6\,\ell_{\rm p}^2\,\sigma^2\,r^6}
\ .
\label{TICFT}
\ee
Hence, for small $|\eta|$, the ratio
\be
^{\rm I}\Gamma_{\rm CFT}\sim \left|
1-{n_{\rm B}\,\ell_{\rm p}^2\,\sigma^2\over
10\,\eta^2\,\pi^2}\right|
\ ,
\ee
which is finite for $\eta\not=0$ and represents a significant
improvement over the BS.
In particular, one has that $^{\rm I}\Gamma_{\rm CFT}\simeq 0$
for
\be
\eta\simeq
\pm {\sqrt{n_{\rm B}}\,\ell_{\rm p}\,\sigma\over\sqrt{10}\,\pi}
\equiv\,^{\rm I}\eta
\ ,
\ee
where we used $\ell_{\rm p}\,\sigma\ll 1$.
For $\eta\simeq\!\,^{\rm I}\eta$ one expects that the holographic
principle yields a reliable description of such black holes.
Note however that the rough estimate $n_{\rm B}\sim {\mathcal N}$ from
Eq.~(\ref{sig}) would yield $|^{\rm I}\eta|\simeq 0.1$
which is significantly larger than the experimental bound
$|\eta|\lesssim 10^{-4}$ from solar system measurements \cite{will}.
\subsubsection{Case II}
From Refs.~\cite{cfm,kar}, let us now consider the metric
described by the functions (for the causal structure see again
Ref.~\cite{cfm})
\be
&&N=\strut\displaystyle{\left[{\eta+\sqrt{1-{2\,M\over r}\,(1+\eta)}
\over 1+\eta}\right]^2}
\nonumber
\\
\label{gII}
\\
&&A=\strut\displaystyle{\left[1-\frac{2\,M}{r}\,(1+\eta)\right]^{-1}}
\ ,
\nonumber
\ee
which yield the non-vanishing Ricci tensor components
\be
^{\rm II}R^r_{\ r}&=&
-2\,^{\rm II}R^\theta_{\ \theta}
=-2\,\,^{\rm II}R^\phi_{\ \phi}
\nonumber
\\
&=&
{2\,\eta\,(1+\eta)\,M\over
\left(\eta+\sqrt{1-{2\,M\over r}\,(1+\eta)}\right)\,r^3}
\ .
\label{RicciII}
\ee
These are again subleading at large $r$ with respect to the
radiation terms in Eq.~(\ref{RicciS}), but of the same order as
those of Case~I, and the estimate in Eq.~(\ref{int}) applies
to this case as well.
The trace of the boson energy-momentum tensor is also of the same
order in $1/r$ as that in Eq.~(\ref{TS}) (see Appendix~\ref{anomalies})
\be
^{\rm II}\expec{T}_{\rm 4D}\simeq
(1+\eta)\,^{\rm S}\expec{T}_{\rm 4D}
\ ,
\label{TII}
\ee
and the conformal anomaly from the AdS-CFT correspondence is
\be
^{\rm II}\expec{T}_{\rm CFT}\sim\,^{\rm I}\expec{T}_{\rm CFT}
\ ,
\label{TIICFT}
\ee
yielding a finite (for $\eta\not=0$) ratio
$^{\rm II}\Gamma_{\rm CFT}\sim\,^{\rm I}\Gamma_{\rm CFT}$ and
$^{\rm II}\Gamma_{\rm CFT}\simeq 0$ for
\be
\eta\simeq
\,^{\rm I}\eta/3
\ ,
\ee
where we again used $\ell_{\rm p}\,\sigma\ll 1$ and $|\eta|\ll 1$.
\subsubsection{Case III}
Finally, from Ref.~\cite{maartens}, the metric
\be
N={1\over A}=1-{2\,M\over r}+{\eta\,M^2\over 2\,r^2}
\ ,
\label{gIII}
\ee
has the Ricci tensor components
\be
^{\rm III}R^t_{\ t}=\,^{\rm III}R^r_{\ r}
=-\,^{\rm III}R^\theta_{\ \theta}
=-\,^{\rm III}R^\phi_{\ \phi}
={\eta\,M^2\over 2\,r^4}
\ .
\label{RicciIII}
\ee
The interval over which such corrections overcome the
Hawking flux is now narrower, namely
\be
1\ll {r\over M}\lesssim\sqrt{|\eta|}\,{M\over\ell_{\rm p}}
\ ,
\label{intIII}
\ee
but still quite large for astrophysical black holes.
The corresponding trace anomaly is given by
\be
^{\rm III}\expec{T}_{\rm 4D}\simeq\,^{\rm S}\expec{T}_{\rm 4D}
\ ,
\label{TIII}
\ee
to leading order in $1/r$ (see Appendix~\ref{anomalies}).
\par
The AdS-CFT trace anomaly is subleading for this case, namely
\be
^{\rm III}\expec{T}_{\rm CFT}\simeq
{\eta^2\,M^4\over \ell_{\rm p}^2\,\sigma^2\,r^8}
\ ,
\label{TIIICFT}
\ee
and
\be
^{\rm III}\Gamma_{\rm CFT}\sim
n_{\rm B}\,{\ell_{\rm p}^2\,\sigma^2\,r^2\over \eta^2\,M^2}
\ ,
\ee
which is larger than those of cases~I and II.
This result seems therefore to disfavor such a metric as a
candidate holographic black hole.
\subsection{Semiclassical stability}
Since the trace $\expec{T}_{\rm 4D}\not=0$ signifies that the
chosen background is not the true vacuum (for which one would
rather expect a vanishing conformal anomaly and toward which
the system will evolve), a quantitative way of estimating the
stability of the above solutions with respect to the BS is
to evaluate the ratio
\be
\Gamma_{\rm 4D}\equiv
\left|{\expec{T}_{\rm 4D}\over ^{\rm S}\expec{T}}\right|
\ .
\ee
In regions where $\Gamma_{\rm 4D}\ll 1$, the corresponding metric
should be more stable than the BS.
This occurs, for instance, for the candidate small black hole
metrics which we shall analyze in Section~\ref{stat_small}
[see Eqs.~(\ref{G4Dsmall})].
Such brane metrics violate the condition (\ref{cut-offs}) and
are therefore unlikely to admit an holographic description, as
our approach will indeed confirm~\footnote{The general
criterion introduced in Section~\ref{general} leads to this
expected property [see Eqs.~(\ref{G2}) and (\ref{Gn})].
This result, in turn, supports the validity of our approach.}.
\subsubsection{Cases I, II and III}
It is interesting to take note of the approximate asymptotic
values of the ratio $\Gamma_{\rm 4D}$ at large $r$ for the three
cases previously discussed:
\be
&&
^{\rm I}\Gamma_{\rm 4D}\to 1+{\eta\over 3}+{\eta^2\over 24}
\nonumber
\\
&&
\nonumber
\\
&&
^{\rm II}\Gamma_{\rm 4D}\to 1+\eta
\label{gaI}
\\
&&
\nonumber
\\
&&
^{\rm III}\Gamma_{\rm 4D}\to 1
\ .
\nonumber
\ee
From such expressions, it appears that the preferred solutions
are again given by cases~I and II, but with $\eta<0$
(in Refs.~\cite{cfm,cm} we already discussed some reasons why
one expects $\eta<0$ and the above results further support this
conclusion since
$\eta>0$ always leads to a larger value of the trace than
$\eta\le 0$).
Case~III instead represents no real improvement over the BS.
\par
\begin{figure}[t]
\centering
\raisebox{4cm}{$\Gamma_{\rm 4D}$}
\epsfxsize=3in
\epsfbox{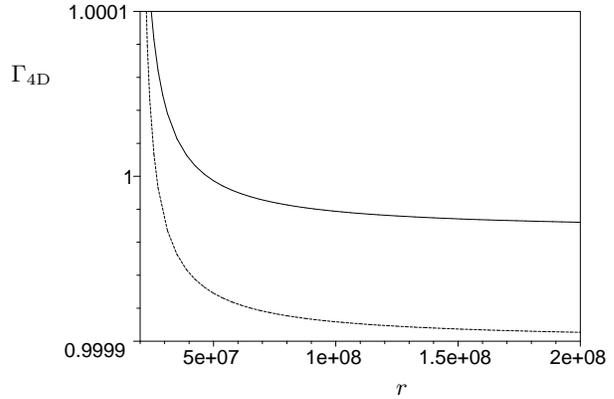}
\\
\raisebox{0.5cm}
{\hspace{2cm} $r$}
\caption{The ratios $^{\rm I}\Gamma_{\rm 4D}$ (solid line) and
$^{\rm II}\Gamma_{\rm 4D}$ (dashed line) for $\eta=-10^{-4}$
and $M=10^7\,\sigma^{-1}$ ($r$ is in units of $\sigma^{-1}$).
\label{gammaI}}
\end{figure}

In details, the ratios $^{\rm I}\Gamma_{\rm 4D}$ and
$^{\rm II}\Gamma_{\rm 4D}$ are plotted in Fig.~\ref{gammaI}
for the typical values
$M=10^7\,\sigma^{-1}\simeq 1\,$km and $\eta=-10^{-4}$ \cite{cm}.
Except for the region very near the horizon ($r_{\rm h}=2\,M$),
the ratios $^{\rm II}\Gamma_{\rm 4D}<\,^{\rm I}\Gamma_{\rm 4D}<1$.
This might signal a stronger instability near the horizon than for
the BS which, however, becomes milder at larger distances.
\subsubsection{Small black holes}
\label{stat_small}
There are more candidate metrics for small black holes
%~\footnote{This
%case violates the condition (\ref{cut-offs}) and the results in this
%Section will further show the failure of the holographic picture.}
with $M\,\sigma\lesssim 1$
(see, e.g., Refs.~\cite{argyres,tanaka,katz}),
namely the higher-dimensional Schwarzschild metrics \cite{myers}
\be
N={1\over A}=1-\left({r_{\rm h}\over r}\right)^n
\ ,
\label{g5D}
\ee
with $n\ge 2$.
Unfortunately one cannot rigorously identify $r_{\rm h}\simeq 2\,M$,
since the four-dimensional ADM mass for this background
is zero~\footnote{This is the reason why such a metric cannot
describe astrophysical black holes \cite{ch}.},
and confronting with the BS (corresponding to $n=1$) becomes more
subtle.
Let us anyways assume $r_{\rm h}\sim M_{(n)}$ holds from Newtonian arguments
(at least for $n=2$ \cite{add,argyres}), where the $M_{(n)}$'s are now
understood as the multipole moments of the energy distribution of the
source.
One then obtains
\be
^{(n)}R^t_{\ t}&=&
\,^{(n)}R^r_{\ r}
=-{n\over 2}\,^{(n)}R^\theta_{\ \theta}
=-{n\over 2}\,^{(n)}R^\phi_{\ \phi}
\nonumber
\\
&=&
{n\,(n-1)\over 2}\,{M^n_{(n)}\over r^{2+n}}
\ .
\label{Ricci5D}
\ee
Note that the scalar
\be
^{(n)}R=(n-1)\,(n-2)\,{M_{(n)}^n\over r^{2+n}}
\ee
just vanishes for $n=1$ (four-dimensional
Schwarz\-schild) and $n=2$ (five-dimensional Schwarz\-schild).
These are the only cases which satisfy the vacuum Eq.~(\ref{R}).
\par
The complete expression for the trace anomaly is given in
Eq.~(\ref{Tn}) which shows that, for $n=2$, one has
%\begin{subequations}
\be
^{(2)}\expec{T}_{\rm 4D}=
n_{\rm B}\,{13\,M^{4}_{(2)}
\over 720\,\pi^2\,r^{8}}
\ ,
\label{T2}
\ee
while, for $n>2$, the leading behavior at large $r$
is given by the $\Box R$ term, that is
\be
^{(n>2)}\expec{T}_{\rm 4D}\sim
n_{\rm B}\,{(n^2-1)\,(n^2-4)\,M^{n}_{(n)}
\over 5760\,\pi^2\,r^{4+n}}
\ ,
\label{Tngtr2}
\ee
%\end{subequations}
both of which never vanish.
However, and although the numerical coefficient is questionable
because of the aforementioned ambiguity in relating $M_{(n)}$
to $r_{\rm h}$, the ratios
%\begin{subequations}
\be
&&^{(2)}\Gamma_{\rm 4D}
\sim\left({M_{(2)}\over r}\right)^2
\nonumber
\\
\label{G4Dsmall}
\\
&&^{(n>2)}\Gamma_{\rm 4D}
\sim\left({M_{(n)}\over r}\right)^{n-2}
\ ,
\nonumber
\ee
%\end{subequations}
tend to zero at large $r$ and are less than one for $n>1$ and
$r\gtrsim M_{(n)}$.
%(see Fig.~\ref{gamma5D} for the cases $n=2$ and $3$).
This makes such metrics better candidates to describe very small
black holes on the brane.
\begin{figure}
\centering
\raisebox{4cm}{$\rho^2$}
\epsfxsize=3in
\epsfbox{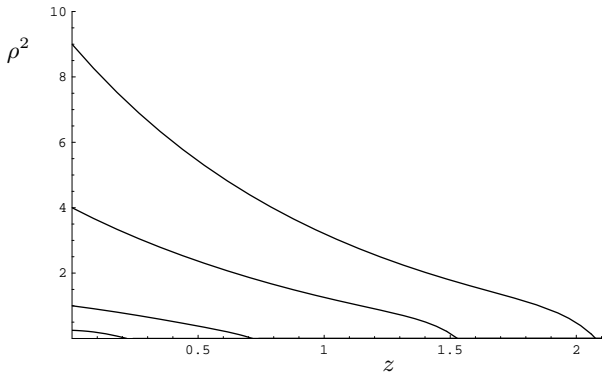}
\\
\raisebox{0.5cm}
{\hspace{2cm} $z$}
\caption{The function $\rho^2$ for $M_{(2)}=\sigma^{-1}$
and $r=M_{(2)}/2$, $M_{(2)}$, $2\,M_{(2)}$ and $3\,M_{(2)}$.
Terms up to order $1/r^{13}$ are included.
\label{R5D}}
\end{figure}
\par
It is then interesting to study their extension in the bulk by
applying the method of Ref.~\cite{cm}.
Details are given in Appendix~\ref{small} for $n=2$
(see also Ref.~\cite{suranyi}).
For $M_{(2)}\,\sigma=1$ the value of $\rho^2$ along geodesic lines of
constant $r$ is displayed in Fig.~\ref{R5D} and the horizon is
approximately given by the line $r=M_{(2)}$.
Note that it is slightly flattened since the maximum value of $z$
along such a line is about $0.7\,\sigma^{-1}$.
It would however depart more and more from that curve the smaller
$M_{(2)}\,\sigma$ is.
For larger values of $M\,\sigma$, one expects a non-vanishing ADM
mass ($2\,M=M_{(1)}$), and the line $r=r_{\rm h}$ is then flatter
and a better approximation of the true location for the horizon
(see Ref.~\cite{cm} for the cases~I, II and III).
\par
The trace anomaly from the AdS-CFT correspondence is given in
Eq.~(\ref{TnCFT}), and just vanishes for $n=1$.
Neglecting numerical coefficients, one thus obtains, for $n=2$,
a ratio
\be
^{(2)}\Gamma_{\rm CFT}\simeq
1-n_{\rm B}\,{13\,\ell_{\rm p}^2\,\sigma^2
\over 720\,\pi^2}
\ ,
\label{G2}
\ee
which vanishes for
\be
n_{\rm B}\simeq{720\,\pi^2\over 13\,\ell_{\rm p}^2\,\sigma^2}
\gg 1
\ ,
\ee
where the inequality follows from the condition (\ref{sig}).
For such a number of boson fields one has
\be
^{(2)}\expec{T}_{\rm 4D}\simeq
{M_{(2)}^4\over\ell_{\rm p}^2\,\sigma^2\,r^8}
\ .
\ee
For $n>2$, instead
\be
^{(n>2)}\Gamma_{\rm CFT}\sim
\left({r\over M_{(n)}}\right)^n
\ ,
\label{Gn}
\ee
which is an increasing function of $n$ and diverges for
$r\to\infty$ as occurred for case~III.
It thus seems that, although such brane metrics are semiclassically
more stable, they significantly depart from the holographic
description for increasing $n$.
This is not contradictory, since the condition (\ref{cut-offs})
[or, equivalently, Eq.~(\ref{large})] is now violated and one
expects that the CFT description fails.
Moreover, one also expects that the smaller the black hole (horizon),
the finer the space-time structure is probed, and one eventually
needs to include stringy effects.
\section{Time-dependent black holes: a conjecture}
\label{time}
Solving Einstein equations for time-dependent metrics
is in general a formidable task, unless symmetries are imposed
to freeze enough degrees of freedom, and there is little hope to
find analytic solutions for the present case~\footnote{There
is an interesting exception: the metric (\ref{gII}) has $R=0$
for $\eta$ and $M$ arbitrary functions of the time (we thank
S.~Kar for pointing this out to us).}.
Let us then begin with a qualitative remark based on the
results we have shown in the previous Section:
Just looking at the trace anomaly (\ref{TS}) one is led to
conclude that the natural time evolution of a four-dimensional
black hole is toward smaller and smaller ADM masses, since
$^{\rm S}\expec{T}_{\rm 4D}=0$ for $M=0$.
However, such an evolution seems to make the black hole less
and less stable, its specific heat being more and more negative
and the temperature diverging, as is argued in the usual canonical
picture of the Hawking radiation \cite{hawking}.
But the picture might change if one considers a fully dynamical
description (for a recent four-dimensional analysis of Hawking
evaporation and trace anomaly, see Ref.~\cite{vilasi}).
\par
In order to substantiate our argument, let us replace
Eq.~(\ref{R}) with the semiclassical brane equation
\be
R=-\ell_{\rm p}^2\,\expec{T}_{\rm 4D}
\ ,
\label{sR}
\ee
and note that, for a metric of the form (\ref{g5D}) and time-independent
$r_{\rm h}$, the scalar curvature $R$ always falls off more slowly
at large $r$ than the corresponding trace anomaly (\ref{Tn}).
Hence, Eq.~(\ref{sR}) cannot be solved by such an {\em ansatz\/}.
However, the situation changes when the metric is time-dependent:
for an asymptotically flat brane metric, on expanding to leading
order at large $r$, the term $\Box R$ becomes of the same leading
order as $R$ and dominates in the expression of the trace
anomaly~\footnote{The precise coefficient
in front of this term depends on the renormalization scheme.
However, since all other possible contributions to the trace
anomaly would still fall off faster then $R$ at large $r$ and we are
just interested in a qualitative result, we shall assume such a factor
is of order 1.}.
In particular, Eq.~(\ref{sR}) becomes
\be
R\sim n_{\rm B}\,\ell_{\rm p}^2\,\ddot R
\ ,
\ee
where a dot denotes the derivative with respect to $t$.
We then reconsider the Schwarzschild metric (\ref{g0}) with the simple
expression for the ADM mass
\be
M=M_0\,e^{-a\,t}
\ ,
\label{exp}
\ee
where $a>0$ so as to enforce decreasing mass.
The Ricci scalar and the trace anomaly for this metric, to
leading order at large $r$, are given by
%\begin{subequations}
\be
&R=\displaystyle 2\,a^2\,e^{-a\,t}\,{M_0\over r}
+{\mathcal O}\left(e^{-2\,a\,t}\,{M_0^2\over r^2}\right)
&
\label{Rab}
\\
\nonumber
\\
&\!\!\!\!\!\!\!\!\!\!
\expec{T}_{\rm 4D}=
-\displaystyle{n_{\rm B}\,a^4\,e^{-a\,t}\,M_0\over 1440\,\pi^2\,r}
+{\mathcal O}\left(e^{-2\,a\,t}\,{M_0^2\over r^2}\right)
&
\ ,
\label{Tab}
\ee
%\end{subequations}
and Eq.~(\ref{sR}) is thus solved to leading order at large
$r$ for
\be
a={\sqrt{2880}\,\pi\over \sqrt{n_{\rm B}}\,\ell_{\rm p}}
\ .
\ee
The CFT trace anomaly in this case is subleading,
\be
\expec{T}_{\rm CFT}\simeq
{a^4\,e^{-2\,a\,t}\,M_0^2\over 6\,\ell_{\rm p}^2\,\sigma^2\,r^2}
\ ,
\ee
and one then concludes that the AdS-CFT correspondence is wildly
violated.
As we mentioned at the end of Section~\ref{stat_small}, this is
not necessarily a flaw.
\par
Finally, the non-vanishing components of the energy-momentum tensor,
again to leading order at large $r$, are given by
\be
\begin{array}{l}
T^t_{\ r}\simeq
-T^r_{\ t}\simeq
\displaystyle{2\,a\,e^{-a\,t}\,M\over \ell_{\rm p}^2\,r^2}
\\
\\
T^\theta_{\ \theta}\simeq
T^\phi_{\ \phi}\simeq
\displaystyle{a^3\,e^{-a\,t}\,M\over \ell_{\rm p}^2\,r}
\ ,
\end{array}
\ee
and the luminosity is
%\begin{subequations}
\be
{\dot M}\sim -a\,M_0\,e^{-a\,t}
\ .
\label{L}
\ee
In order to fix a reasonable value for $M_0$, we can now assume
that, for sufficiently large ADM mass, the standard Hawking relation
holds \cite{hawking},
\be
{\dot M}\sim -n_{\rm B}\,K\,{\ell_{\rm p}^2\over M^2}
\ ,
\label{LH}
\ee
%\end{subequations}
where $K$ is the same coefficient which appears in Eq.~(\ref{T_S}).
The transition to the new regime would then occur when the two
expression of the luminosity, Eqs.~(\ref{L}) and (\ref{LH}), match
(at $t=0$), that is for
\be
M_0&\sim&\left(n_{\rm B}\,{\ell_{\rm p}^2\over a}\right)^{1/3}
\simeq
0.1\,n_{\rm B}^{1/2}\,\ell_{\rm p}
\nonumber
\\
&\simeq& 0.1\,\sigma^{-1}
\lesssim 0.1\,{\rm mm}
\ ,
\ee
where we have estimated $n_{\rm B}$ as in Eq.~(\ref{sig}) in the
second line.
Since $\sigma\,M_0<1$, it is no more a surprise that the
holographic description fails for the present case.
Note that the luminosity (\ref{L}) vanishes for $M=0$ [whereas the
expression in Eq.~(\ref{LH}) diverges], that is the temperature of
such black holes is much lower than the canonical one.
This is just the kind of improvement one expects from energy conservation
and the use of the microcanonical picture for the system consisting
of the black hole and its Hawking radiation \cite{harms} near the
end-point of the evaporation \cite{mfd}~\footnote{Analogous results
have been obtained in two dimensions \cite{easson} for dilatonic
black holes which satisfy a principle of least
curvature \cite{trodden}.}.
\par
Of course, the above calculations are just suggestive of how to tackle
the problem, and are not meant to be conclusive.
One point is however clear, that in a brane-world scenario one has
to accommodate just for the one vacuum condition in Eq.~(\ref{sR}),
which is therefore easier to approach than the four-dimensional
analogue.
The hard part of the task is then moved to the bulk:
the brane metric we have considered must not give rise to spurious
singularities off the brane.
If the evaporation is complete, this is obviously true for the
Schwarzschild metric and (\ref{exp}) in the limit $t\to \infty$,
but a complete analysis of the bulk equations for such a
time-dependent brane metric is intractable analytically.
\par
Let us finally speculate on the basis that a vanishing four-dimensional
ADM mass is not equivalent to zero proper mass, since terms of higher
order in $1/r$, such as those considered in Eq.~(\ref{g5D}), may
survive after the time when $M$ has vanished (or, rather, approached
the critical value $\ell_{\rm g}$).
They are in general expected to appear as generated by the
non-vanishing $E_{ij}$ and the trace anomaly (\ref{Tn}) they
give rise to is smaller for larger $n$ (and the same value of the
``mass'' parameter $M_{(n)}$).
This opens up a wealth of new possibilities for the brane-world.
Since we have shown evidence that the late stage of the evaporation
is likely a (slow) exponential decay, one can capture an instantaneous
picture of its evolution in time [i.e., apply the adiabatic
approximation in order to obtain the static form (\ref{g})] and
expand that in powers of $1/r$,
\be
N=1-\sum_{n=\bar n}\left({M_{(n)}\over r}\right)^n
\ ,
\ee
where $M_{(1)}\equiv 2\,M$ and $n=\bar n$ is the smallest order
for which the coefficient $M_{(n)}\not=0$, so that, although the
black hole remains five-dimensional, its profile looks like it
is higher-dimensional.
Then, one would also have a ``remnant'' trace anomaly which is
approximately given by the expression in Eq.~(\ref{Tn}) with
$n=\bar n$.
If $\bar n$ increases in time, the corresponding trace anomaly
decreases in time and the black hole appears as a higher and
higher dimensional object from the point of view of an observer
restricted on the four-dimensional brane-world.
Correspondingly, the space-time (brane) around the singularity
looks flatter and flatter [see the Ricci tensor elements in
Eq.~(\ref{Ricci5D})].
More precisely, once the horizon radius has approached $\ell_{\rm g}$,
a geometric description of the space surrounding the central
singularity becomes questionable, and just the large $r$ limit
of the metric can be given sense.
The latter is practically flat for $r>M_{(\bar n)}$ when
$\bar n\ge 2$.
\begin{acknowledgments}
I thank G.~Alberghi, F.~Bastianelli, C.~Germani, B.~Harms and
L.~Mazzacurati for comments and suggestions.
\end{acknowledgments}
\appendix
\section{Trace anomalies}
\label{anomalies}
We display here the complete expressions of the trace anomalies
for the static cases I, II and III of Section~\ref{static}, and
$n_{\rm B}=1$.
In obvious notation:
\begin{widetext}
\begin{subequations}
\be
^{\rm I}\expec{T}_{\rm 4D}
&=&{\left(1-{3\,M\over 2\,r}\right)^{-4}M^2
\over 25920\,\pi^2\,r^6}
\left[18\,(24+8\,\eta+\eta^2)
-16\,(162+81\,\eta+10\,\eta^2)\,{M\over r}
\right.
\nonumber
\\
&&\left.
%\phantom{{\left(1-{3\,M\over 2\,r}\right)^{-4}M^2
%\over 25920\,\pi^2\,r^6}\ }
+12\,(486+315\,\eta+49\,\eta^2)\,{M^2\over r^2}
+216\,(27+21\,\eta+4\,\eta^2)\,{M^3\over r^3}
%\right.
%\nonumber
%\\
%&&\left.
%\phantom{{\left(1-{3\,M\over 2\,r}\right)^{-4}M^2
%\over 25920\,\pi^2\,r^6}\ }
+9\,(243+216\,\eta+48\,\eta^2)\,{M^4\over r^4}
\right]
\ ,
\ee
\be
^{\rm II}\expec{T}_{\rm 4D}
&=&{(1+\eta)^2 M^2\over 240\,\pi^2\,r^6}\,
\left(\eta+\sqrt{1-{2\,M\over r}\,(1+\eta)}\right)^{-4}
%\nonumber
%\\
%&&\times
\left[4+{3\over 2}\,\eta^2\,(9+\eta^2)+\eta\,(12+7\,\eta)\,
\sqrt{1-{2\,M\over r}\,(1+\eta)}
\right.
\nonumber
\\
&&\left.
\phantom{\times[]}
-(1+\eta)\,\left(16+27\,\eta^2+24\,\eta\,
\sqrt{1-{2\,M\over r}\,(1+\eta)}\right)\,{M\over r}
%\right.
%\nonumber
%\\
%&&\left.
\phantom{\times[]}
+16\,(1+\eta)^2\,{M^2\over r^2}
\right]
\ ,
\ee
\be
^{\rm III}\expec{T}_{\rm 4D}
&=&{M^2\over 60\,\pi^2\,r^6}\,\left(
1-\eta\,{M\over r}+{13\,\eta^2\,M^2\over 48\,r^2}
\right)
\ ,
\ee
\be
^{(n)}\expec{T}_{\rm 4D}=
{M^{n}_{(n)}\over 5760\,\pi^2\,r^{4+n}}\,
\left[(n^2-1)\,(n^2-4)
-(3\,n^4-8\,n^3-23\,n^2+4)\,{M^n_{(n)}\over r^n}
\right]
\ .
\label{Tn}
\ee
\end{subequations}
From the AdS-CFT correspondence \cite{skenderis} one instead
obtains
\begin{subequations}
\be
^{\rm I}\expec{T}_{\rm CFT}=
{\eta^2\,M^2\over 6\,\ell_{\rm p}^2\,\sigma^2\,r^6}\,
\left(1-{3\,M\over 2\,r}\right)^{-4}\,
\left(1-{8\,M\over 3\,r}+{2\,M^2\over r^2}\right)
\ ,
\ee
\be
^{\rm II}\expec{T}_{\rm CFT}=
{3\,\eta^2\,(1-\eta)^2\,M^2\over 2\,\ell_{\rm p}^2\,\sigma^2\,r^6}\,
\left(\eta+\sqrt{1-{2\,M\over r}\,(1+\eta)}\right)^{-2}
\ ,
\ee
\be
^{\rm III}\expec{T}_{\rm CFT}=
{\eta^2\,M^2\over 4\,\ell_{\rm p}^2\,\sigma^2\,r^8}
\ ,
\ee
\be
^{(n)}\expec{T}_{\rm CFT}
={n^2+8\,n+4\over 24}\,{(1-n)^2\,M^{2\,n}_{(n)}
\over \ell_{\rm p}^2\,\sigma^2\,r^{2\,(2+n)}}
\ .
\label{TnCFT}
\ee
\end{subequations}
\section{Small black holes}
\label{small}
In the approach of Ref.~\cite{cm} the bulk metric is taken
of the form
\be
\dd s^2=-N(r,z)\,\dd t^2+A(r,z)\,\dd r^2+\rho^2(r,z)\,\dd \Omega^2
+\dd z^2
\ .
\label{gbulk}
\ee
For the brane metric in Eq.~(\ref{g5D}) with $n=2$, the computed
metric components (to order $1/r^6$, for the sake of simplicity)
are then given by
\begin{subequations}
\be
N&=&e^{-\sigma\,z}\left\{1
-{M_{(2)}^2\over r^2}
+\left(1 - e^{\sigma\,z}\right)^2
{M_{(2)}^2\over\sigma^2\,r^4}
\right.
\\
&&\left.\ \ \
-\left[
2 + \sigma^2\,M_{(2)}^2
-2\,e^{\sigma\,z}\left(6 + \sigma^2\,M_{(2)}^2\right)
+e^{2\,\sigma\,z}
\left(6 + \sigma^2\,M_{(2)}^2 - 12\,\sigma\,z\right)
+ 4\,e^{3\,\sigma\,z}
\right]
{M_{(2)}^2\over \sigma^4\,r^6}
\right\}
\nonumber
\\
&&
\nonumber
\\
A&=&e^{-\sigma\,z}\left\{1
+{M_{(2)}^2\over r^2}
+\left[\sigma^2\,M_{(2)}^2+\left(1 - e^{\sigma\,z}\right)^2\right]
{M_{(2)}^2\over \sigma^2\,r^4}
\right.
\\
&&\left.
-\left\{
3\,\left(1 + e^{\sigma\,z}\right)^2\,\sigma^2\,M_{(2)}^2
-3\,\sigma^4\,M_{(2)}^4
+2\,\left[1-6\,e^{\sigma\,z}
+ 3\,e^{2\,\sigma\,z}\,\left(1 - 2\,\sigma\,z\right)
+2\,e^{3\,\sigma\,z}
\right]\right\}
{M_{(2)}^2\over  3\,\sigma^4\,r^6}
\right\}
\nonumber
\\
&&
\nonumber
\\
\rho^2&=&r^2\,e^{-\sigma\,z}\left\{1
-\left(1-e^{\sigma\,z}\right)^2
{M_{(2)}^2\over \sigma^2\,r^4}
+\left[1 - 6\,e^{\sigma\,z}
+3\,e^{2\,\sigma\,z}\,\left(1 - 2\,\sigma\,z\right)
+2\,e^{3\,\sigma\,z}
\right]
{4\,M_{(2)}^2\over 3\,\sigma^4\,r^6}
\right\}
\ .
\ee
\end{subequations}
\end{widetext}
Note that, as we commented upon in Ref.~\cite{cm}, the function
$\rho^2$ vanishes for a finite value of $z$, with $r$ held fixed,
and that locates the axis of cylindrical symmetry in the Gaussian
normal reference frame.
The latter covers the whole bulk manifold of such black holes,
since no crossing occurs between lines of different constant
$r$ (see Fig.~\ref{R5D}).
%
%\end{widetext}
%
%
%
%
%
%

%
%
\end{document}